\documentclass[onecolumn,showpacs,preprintnumbers,amsmath,amssymb,showkeys,11pt]{revtex4}
\usepackage{graphicx}
\usepackage{dcolumn}
\usepackage{bm}

\DeclareMathOperator{\BXOR}{BXOR}
\DeclareMathOperator{\AsymCSS}{AsymCSS} 
\DeclareMathOperator{\W}{W}
\DeclareMathOperator{\BB}{BB84}

\DeclareMathOperator{\ld}{log_2}
\DeclareMathOperator{\id}{id}
\DeclareMathOperator{\at}{artanh}
\newcommand{\ens}[0]{\ensuremath} 
\newcommand{\ket}[1]{\ensuremath{|#1 \rangle}} 
\newcommand{\bra}[1]{\ensuremath{\langle #1|}} 
\newcommand{\pr}[1]{\ens{\ket{#1}\bra{#1}}} 
\newcommand{\x}[0]{\ens{\otimes}} 
\newcommand{\impl}[0]{\ens{\Rightarrow}} 
\newcommand{\eqv}[0]{\ens{\Leftrightarrow}} 
\newcommand{\nach}[0]{\ens{\rightarrow}} 
\newcommand{\nGr}[0]{\ens{{n \rightarrow \infty}}} 
\newcommand{\Bew}[0]{\emph{Proof: }} 
\newcommand{\Mge}[2]{\ens{\left\lbrace #1|\,#2 \right\rbrace}}
\newcommand{\Mg}[1]{\ens{\left\lbrace #1 \right\rbrace}}
\newcommand{\Mgn}[2]{\ens{\Mg{#1,\dots,#2}}} 
\newcommand{\MgN}[1]{\ens{\Mg{0,\dots,#1}}} 
\newcommand{\Fkt}[3]{\ens{#1 : #2 \nach #3}} 

\def\Eins{{\leavevmode{\rm 1\ifmmode\mkern -4.4mu\else\kern -.3em\fi I}}}
\newcommand{\eps}[0]{\ens{\varepsilon}}
\newcommand{\cS}[0]{\ens{\mathcal{S}}}
\newcommand{\N}[0]{\ens{\mathbb{N}}}
\newcommand{\R}[0]{\ens{\mathbb{R}}}
\newcommand{\F}[0]{\ens{\mathbb{F}}}
\newcommand{\Sbd}[0]{\ens{\cS_{\mathrm{bd}}}}
\newcommand{\Sv}[0]{\ens{\cS_{\mathrm{v}}}}
\newcommand{\oSv}[0]{\ens{\overline{\Sv}}}
\newcommand{\lm}[2]{\ens{\left(#1,#2\right)}}
\newcommand{\dd}[0]{\ens{33.\overline{3}}} 
\newcommand{\Bn}[0]{\ens{B_n}}
\newcommand{\Pn}[0]{\ens{P_n}}
\newcommand{\Sn}[0]{\ens{S_n}}

\newtheorem{Definition}{Definition}
\newtheorem{Satz}{Theorem}
\newtheorem{Lemma}{Lemma}
\newtheorem{Korollar}{Corollary}
\newtheorem{Vermutung}{Conjecture}

\begin{document}

\title{Asymptotic correctability of Bell-diagonal quantum states \\and maximum tolerable bit error rates}
\author{Kedar S. Ranade}
\author{Gernot Alber}
\affiliation{Institut f\"ur Angewandte Physik, Technische Universit\"at Darmstadt, 64289 Darmstadt, Germany}
\date{October 5, 2005}

\pacs{03.67.Mn, 03.67.Dd, 03.67.-a}
\keywords{quantum state purification, quantum cryptography, maximum tolerable error rates,
  Gottesman-Lo-type protocols}

\begin{abstract}
The general conditions are discussed which quantum state purification protocols have to fulfill in order
to be capable of purifying Bell-diagonal qubit-pair states, provided they consist of steps that map
Bell-diagonal states to Bell-diagonal states and they finally apply a suitably chosen Calderbank-Shor-Steane
code to the outcome of such steps. As a main result a necessary and a sufficient condition on asymptotic
correctability are presented, which relate this problem to the magnitude of a characteristic exponent
governing the relation between bit and phase errors under the purification steps.
These conditions allow a straightforward determination of maximum tolerable bit error rates
of quantum key distribution protocols whose security analysis can be reduced to the purification of
Bell-diagonal states.
\end{abstract}

\maketitle

\section{Introduction}
The quantum cryptographic protocol developed by Bennett and Brassard (BB84)~\cite{BB} demonstrates in an
impressive way how the key distribution problem of classical cryptography can be solved by means of
quantum physics. Later Shor and Preskill \cite{SP} demonstrated that the security of this quantum key
distribution protocol is guaranteed at least up to bit error rates of approximately $11.4\,\%$.
Their proof is based on two main ideas.
Firstly, it exploits an equivalence between the originally proposed BB84 protocol as a prepare-and-measure
protocol and an associated entanglement-based protocol.
Secondly, it reduces the security issue to the capability of purifying Bell-diagonal qubit-pair states with
the help of one-way classical communication and Calderbank-Shor-Steane (CSS) codes \cite{CSS1,CSS2}.
Gottesman and Lo \cite{GL} extended Shor and Preskill's approach to entanglement purification protocols
which involve bit- and phase-error correcting sequences based on classical two-way communication 
followed by a CSS-based entanglement purification step.
This way they were able to raise the maximum tolerable bit error rate of the BB84 protocol to $18.9\, \%$.
Later on Chau \cite{Ch} extended this approach thereby achieving a maximum tolerable bit error rate of
$20\,\%$.
Motivated by these investigations of Gottesman and Lo in this work general entanglement purification protocols 
are analyzed which imply the security of any quantum key distribution protocol whose security analysis
can be reduced to the purification of Bell-diagonal states. The BB84 protocol and the highly symmetric
six-state protocol \cite{Br} are well-known examples of such quantum key distribution protocols. 
The general entanglement purification protocols considered are supposed to map Bell-diagonal states to
Bell-diagonal states until the Shannon bound guarantees a successful completion of the entanglement
purification on the basis of an appropriate CSS encoding and classical one-way communication.
A special example thereof is the entanglement purification protocol introduced by Gottesman and Lo, which,
in addition, is compatible with a reduction of an entanglement-based quantum key distribution protocol
to an associated prepare-and-measure scheme.
As a main result a necessary and a sufficient condition (main theorem) on asymptotic correctability of
Bell-diagonal qubit-pair states are presented relating the success of such a general entanglement purification
protocol to the magnitude of a characteristic exponent, which governs the scaling between bit and phase errors.
This latter characteristic exponent can be determined in a straightforward way and allows the determination
of maximum tolerable bit error rates of the Bell-diagonal states involved. Applying this general result to
entanglement purification protocols of the Gottesman-Lo type, for example, this criterion implies that even
without any phase-error correcting steps of the Gottesman-Lo type secret keys can be generated by
the BB84 and six-state quantum cryptographic protocols
up to the already known bit error rates of $1/5 = 20\% $ and $1/2-1/(2\sqrt{5})\approx 27.6393\,\%$ \cite{Ch}
and that in the absence of phase-error correction no higher bit error rates are tolerable.
Furthermore, numerical evidence is provided that also arbitrary additional sequences of phase-error
correcting steps cannot improve on these particular bounds.

\par This manuscript is organized as follows: In order to put the general entanglement purification protocols
considered in our main theorem into perspective  we first of all summarize basic aspects of the entanglement
purification protocol of Gottesman and Lo \cite{GL} and generalize their original proposal to arbitrary
numbers $n$ of qubit pairs. Correspondingly, basic notions together with the generalized
bit-error (\Bn) and phase-error (\Pn) correcting Gottesman-Lo-type steps are introduced in section~2.  
In section~3 basic asymptotic properties of these purification steps are analyzed for large numbers of
qubit pairs. In particular, the exponents characterizing the scaling of the bit and phase errors under
\Bn\, and \Pn\, steps are determined.
Our main theorem concerning the asymptotic correctability of entanglement purification of Bell-diagonal
states and its relation to the exponents characterizing bit and phase errors is stated and proved in section~4.
Finally, based on this main theorem in section~5 the asymptotic correctability of the \Bn\, and \Pn\, steps
characterizing Gottesman-Lo-type purification protocols are investigated in more detail.
It is shown that bit-error correcting \Bn\, steps alone are already able to guarantee security of the
BB84 protocol and the six-state protocol up to maximum bit error rates of magnitude $1/5$ and
$1/2-1/(2\sqrt{5})$, respectively. Furthermore, numerical evidence is provided that even arbitrary
sequences of phase-error correcting \Pn\, steps cannot improve on these bounds. Based on this evidence these
numbers constitute the maximum possible error rates which are tolerable in the BB84 protocol and in the
six-state protocol provided error correction and privacy amplification are based on arbitrary sequences of
\Bn\, and \Pn\, steps of the Gottesman-Lo type. For the sake of a clearer presentation of the main ideas
some proofs of theorems stated in these sections are postponed to the appendices. A more detailed elaboration
of some statements can be found in \cite{KSR}.

\section{Purification protocols of the Gottesman-Lo type}
In this section basic properties of bit-error (\Bn) and phase-error (\Pn) correction steps are discussed
which generalize the bit- and phase-error correcting steps $B_\mathrm{GL}$ and $P_\mathrm{GL}$
proposed by Gottesman and Lo \cite{GL} to arbitrary numbers $n$ of qubit pairs.
These steps are capable of reducing the bit and phase errors of Bell-diagonal qubit-pair states and can
be used as  building blocks of entanglement purification protocols which are based on classical two-way
communication. In view of the Gottesman-Lo theorem \cite{GL}
entanglement purification protocols consisting of these \Bn\, and \Pn\, steps can be reduced to
prepare-and-measure schemes.
\par Gottesman and Lo proved that it is sufficient for guaranteeing security of the BB84 and the six-state
protocol to be able to purify classical mixtures of the four (pure) Bell states 
\begin{equation}\begin{array}{rlrlrlrl}
  \ket{\Phi^\pm} &:= (1/\sqrt{2})\bigl[\ket{00}\pm\ket{11}\bigr], \qquad
  \ket{\Psi^\pm} &:= (1/\sqrt{2})\bigl[\ket{01}\pm\ket{10}\bigr].
\end{array}\end{equation}
If necessary, the following notation will be used \cite{BDSW}: $(0,0) := \ket{\Phi^+}$,
$(1,0) := \ket{\Phi^-}$,
$(0,1) := \ket{\Psi^+}$, $(1,1) := \ket{\Psi^-}$. Here, the numbers are to be understood as elements
of the binary field $\F_2$. Mixtures of Bell states are denoted by
\begin{equation}
  (a,b,c,d) := a\,\pr{\Phi^+} + b\,\pr{\Phi^-} + c\,\pr{\Psi^+} + d\,\pr{\Psi^-}
\end{equation}
with $a,b,c,d \geq 0$ and $a+b+c+d = 1$. The set of all such Bell-diagonal states is denoted
by $\Sbd$. A Bell-diagonal state is entangled, if and only if one of the four
coefficients is larger than $1/2$ \cite{BDSW}. In our discussion a Bell-diagonal state will be called
entangled with respect to \ket{\Phi^+}, if $a > 1/2$. The set of states with $a > 1/2$ and with $a \geq 1/2$
are denoted by \Sv\, and by \oSv, respectively.
\par In the subsequent discussion we choose the state $\ket{\Phi^+}$ as the reference state for entanglement
purification; therefore $a \equiv F$ will be called fidelity (with respect to \ket{\Phi^+}). Furthermore,
the parameters  $b$, $c$, and $d$ are the pure phase error rate, the pure bit error rate and the combined
bit-phase error rate. Correspondingly, the parameters $B = c+d$ and $P = b+d$ are the total bit and phase
error rates.
\par For the purposes of entanglement purification it is sufficient to assume that Alice and Bob share
an infinite number of qubit pairs, all described by the same density operator $\rho = (a,b,c,d) \in \Sv$
\cite{GL,LC,X-BW}. All purification steps considered act as mappings on the set $\Sbd$. A particular step
of the purification protocols considered takes a fixed number $n$ of qubit pairs, all prepared in
the same state $\rho = (a,b,c,d)$, as input and yields with some non-vanishing probability, which may
depend upon~$\rho$, a final qubit pair in the state $\rho' = (a',b',c',d')$ or no qubit pair at all.

\subsection{\Bn\,steps}
A \Bn\, step which involves $n \in \N$ qubit pairs reduces the bit error rate, but simultaneously it also
increases the phase error rate of the original quantum state. 
It is defined by the following sequence of steps:
\begin{enumerate}
  \item Alice and Bob choose $n$ qubit pairs $QP_1, \dots, QP_n$.
  \item Alice and Bob apply bilateral $\BXOR$ operations of the form $\BXOR(QP_1,QP_k)$ for all qubit pairs
    $k \in \Mg{2,\dots,n}$ ($n-1$ operations).
  \item Alice and Bob measure the bit parities of all pairs from $QP_2$ to $QP_n$ and continue using $QP_1$,
    if and only if all parities are $+1$ (same bit values for Alices and Bobs measurement).
    The pairs $QP_2,\dots,QP_n$ are discarded.
\end{enumerate}
Here, the $\BXOR$ operation on Bell-diagonal states is defined by \cite{GL, BDSW}
\begin{equation}
  \BXOR(QP_1,QP_2) : (l_1,m_1) \x (l_2,m_2) \mapsto (l_1 \oplus l_2, m_1) \x (l_2,m_1\oplus m_2).
\end{equation}
Thus, for a given set of $n$ pure Bell pairs $\lm{l_i}{m_i}$, according to step (ii) the $\BXOR$ operations
are equivalent to the transformation
\begin{equation}
  \bigotimes\nolimits_{i = 1}^{n}\,\,\lm{l_i}{m_i} \mapsto \lm{\bigoplus\nolimits_{i = 1}^{n} l_i}{m_1}
    \x \left[\bigotimes\nolimits_{k = 2}^{n} \lm{l_k}{m_1 \oplus m_k}\right].
\end{equation}
According to step (iii) the pair $QP_1$ is kept for  the next step, if  $m_1 \oplus m_k = 0$ holds for
all $k \in \Mg{2,\dots,n}$. Otherwise this qubit pair is discarded. Therefore, we obtain the relations
$B_1 = \id_{\Sbd}$, $B_2 = B_{\mathrm{GL}}$, $\Bn B_m = B_{nm}$, and $(B_{\mathrm{GL}})^n = B_{2^n}$.
\par Note that Alice and Bob could perform the measurements of the pairs $QP_2, \dots, QP_n$ immediately after
the respective $\BXOR$ operation. If the pair $QP_1$ is discarded immediately after the first false parity,
the average number of discarded qubits reduces, which results in a higher key generation rate.
\par In \ref{AnhBn} it is shown that with respect to the first qubit pair $QP_1$ a \Bn\, step can be
identified with a mapping \Fkt{\Bn}{\Sbd}{\Sbd} with $\Bn : (a,b,c,d) \mapsto (a',b',c',d')$ and with
\begin{equation}\begin{array}{rlrlrlrl}\label{BnA}
  a' &= \bigl[(a+b)^n + (a-b)^n\bigr] / 2N, \qquad
  b' &= \bigl[(a+b)^n - (a-b)^n\bigr] / 2N,\\
  c' &= \bigl[(c+d)^n + (c-d)^n\bigr] / 2N, \qquad
  d' &= \bigl[(c+d)^n - (c-d)^n\bigr] / 2N.
\end{array}\end{equation}
The value $N = \bigl[(a+b)^n + (c+d)^n\bigr] $ is the survival probability of the first pair.

\subsection{\Pn\,steps}
In analogy to the $B_{\mathrm{GL}}$\ step also the \Bn\, step can be adapted to correct phase errors~\cite{GL}.
However, according to the Gottesman-Lo theorem such a step has the disadvantage that it cannot be reduced
to some prepare-and-measure protocol. Therefore, Gottesman and Lo originally developed an alternative
phase-error correction step which is not as efficient, but which can be reduced to a prepare-and-measure
protocol. The \Pn~step considered in the following is a generalization of this step originally developed
by Gottesman and Lo \cite{GL}. For any $n \in \N_0$, we define a $P_{2n+1}$ step as follows:
\begin{enumerate}
  \item Alice and Bob choose $2n+1$ qubit pairs $QP_1, \dots, QP_{2n+1}$.
  \item Alice and Bob perform Hadamard transformations \cite{GL,NC} on all pairs.
  \item Alice and Bob perform $\BXOR$ operations of the form $\BXOR(QP_1,QP_k)$ for all
    qubit pairs with $k \in \Mgn{2}{2n+1}$ ($2n$ operations).
  \item Alice and Bob measure the bit parities of all pairs from $QP_2$ to $QP_n$; the number of pairs
    with bit parity $-1$ (different outcomes for Alice and Bob) is denoted as $m \in \MgN{2n}$.
  \item Alice and Bob perform a Hadamard transformation on $QP_1$.
  \item If $m \geq n + 1$, Bob performs the transformation $\Eins \x \sigma_z$ on the first pair.
    Otherwise, Bob leaves the first pair unchanged. The pairs $QP_2,\dots, QP_{2n+1}$ are discarded.
\end{enumerate}
If in step (v) Alice and Bob apply the Hadamard transformation to all qubit pairs, they can exchange
steps (iv) and (v), if they measure the phase parity $l_1\,\oplus\,l_k$ instead of the bit parity for
$k \in \Mg{2,\dots,2n+1}$. In this latter case the transformation yields
\begin{equation}\label{PnTransf}
  \bigotimes\nolimits_{i = 1}^{2n+1}\,\,\lm{l_i}{m_i} \mapsto \lm{l_1}{\bigoplus\nolimits_{i = 1}^{2n+1} m_i}
   \x \left[\bigotimes\nolimits_{k = 2}^{2n+1} \lm{l_1 \oplus l_k}{m_k}\right].
\end{equation}
According to Bob's final transformation in step (vi) the new phase of the first qubit pair $QP_1$,
as characterized by the parameter $l_1$, is fixed by the majority of the $2n+1$ phases of all
qubit pairs involved.
\par Similar to the case of the \Bn\, step, we obtain $P_1 = \id_{\Sbd}$ and $P_3 = P_\mathrm{GL}$.
But contrary to the case of \Bn\, steps, a sequence $P_nP_m$ is always worse than a single $P_{nm}$\,
step. This originates from the fact that the bit errors introduced by $\Pn P_m$ and $P_{nm}$ sequences
are always equal, whereas the majority of majorities is not necessarily the total majority of phases. Note that
the use of a \Pn\, step is equivalent to the application of the $[n,1,n]$ code in~\cite{Ch}.
\par Calculating the evolution resulting from the application of a \Pn\, step is much more complicated
than the resulting evolution of \Bn\, steps as given in (\ref{BnA}). However, it turns out that the
evolution of bit and phase errors $B$ and $P$ can be determined easily (compare with (\ref{PnUngl})).

\subsection{Remarks}
Note that the bit error rates after applying \Bn\, or \Pn\, steps depend only on the
previous bit error rate (but not on the phase error rate); similarly, the new phase error rate after using a
\Pn\, step depends only on the previous phase error rate. Using \Bn\, steps, the exact coefficients
determine the evolution of the phase error rate; considering $\rho \in \Sv$ and \nGr, the evolution
is mostly determined by the fidelity~$a$ and the pure phase error rate~$b$.
\par In particular, when using \Bn\, and \Pn\, steps only, Alice and Bob do not gain any advantage,
if they measure bit errors after performing some of these steps. This seems to be obvious considering
the fact that they can be reduced to prepare-and-measure-schemes, where phase errors cannot have any
influence on the protocol.

\section{Asymptotic evolution of \Bn\, and \Pn\, steps}
In this section the evolution of Bell-diagonal qubit-pair states is investigated, if they are subjected to
\Bn\, and \Pn\, steps. Here, the asymptotic evolution for large values of $n$ is of particular interest.
In the subsequent discussion this asymptotic evolution is characterized by exponents $r$ and $r_P$ for
\Bn\, and \Pn\, steps, respectively, which determine the relative scaling between bit and
phase errors. As demonstrated in detail in section~\ref{KorrKrit} the values of these
characteristic exponents are directly related to the correctability of Bell-diagonal quantum states.

\subsection{Asymptotic evolution of \Bn\, steps}\label{AsymBn}
\begin{figure}[t]\begin{center}
  \includegraphics[height=145pt,width=145pt]{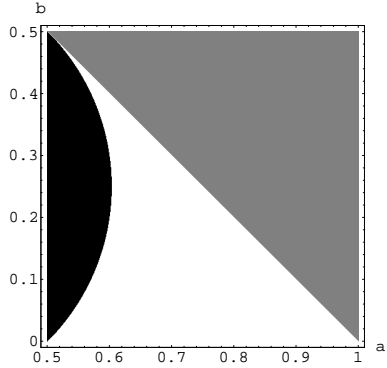}
  \caption{Regions for $r > 2$ (white) and $r \leq 2$ (black) for fidelity $a$ and pure phase error rate
    $b$; grey: no physical states.}
  \label{rBer}
\end{center}\end{figure}

Let us consider the evolution of the quantum state $\rho = (a,b,c,d) \in \Sv$ of a single qubit pair using
\Bn\, steps for large values of $n$. For the sake of simplicity it is assumed that $b > 0$ and $c + d > 0$,
because the remaining cases are trivial. For this purpose we define first of all some useful variables: 
\begin{equation}
  \tilde x := \frac{a+b}{c+d}, \qquad \Delta_1 := \frac{a-b}{c+d}, \qquad \Delta_2 := \frac{c-d}{c+d}.
\end{equation}
After having performed a \Bn\, step the resulting quantum state is given by
\begin{equation}\label{xnyn}
(a',b',c',d') \equiv
  \left(\frac{1}{2} - x_n + y_n + \delta_n, \frac{1}{2} - y_n - \delta_n, x_n - \delta_n, \delta_n\right)
    := \Bn\Bigl[(a,b,c,d)\Bigr],
\end{equation}
where $x_n$ and $y_n$ denote the resulting \emph{bit error rate} ($B$) and \emph{inverse phase error rate}
($1/2-P$). The quantity $\delta_n$ characterizes a correlation between bit and phase errors.
The evolution (\ref{BnA}) immediately implies (the symbol $\doteq$ means asymptotically equal)
\begin{equation}\begin{array}{rlrlrlrl}
  x_n  &= (1 + \tilde x^n)^{-1}                      && \doteq \tilde x^{-n},\\
  2y_n &= (\Delta_1^n + \Delta_2^n)/(1 + \tilde x^n) &&\doteq \Delta_1^n/\tilde x^n.
\end{array}\end{equation}
For particular values of the parameters $a, b, c, d$ it is possible to define a characteristic exponent
$r \in \R$  with the defining property $\lim_\nGr x_n/(2y_n)^r = 1$. In view of the elementary relation
\begin{equation}\label{BnKonv}
  \frac{x_n^{1/r}}{2y_n} = \frac{(1 + \tilde x^n)^{1-1/r}}{\Delta_1^n + \Delta_2^n}
    \doteq \frac{\tilde x^{n(1-1/r)}}{\Delta_1^n}
   = \left(\frac{\tilde x^{1-1/r}}{\Delta_1}\right)^n,
\end{equation}
this defining property implies that the term in the bracket must be unity, i.\,e.
\begin{equation}
  \tilde x^{(1-1/r)} = \Delta_1 \eqv r = \left[1 - \frac{\ln \Delta_1}{\ln \tilde x}\right]^{-1}
    = \frac{\ln \frac{a+b}{c+d}}{\ln \frac{a+b}{a-b}}.
\end{equation}
Therefore, using the conservation of probability, i.\,e. $c + d = 1 - a - b$, one may establish relations
between values of the characteristic parameter $r$ and particular Bell-diagonal states. Two examples
of such correlations are:
\begin{equation}\begin{array}{rlrlrlrl}
  r > 1 &\eqv a > 1/2\,\mathrm{(entanglement\,w.r.t.\,\ket{\Phi^+})},\\
  r > 2 &\eqv f(a,b) := a^2 + b^2 - (a+b)/2 > 0\label{fFkt}\\
        &\eqv \left(a-1/4\right)^2 + \left(b-1/4\right)^2
          > \left(1/2\sqrt{2}\right)^2 = 1/8.
\end{array}\end{equation}
The left hand side of the latter inequality can be interpreted geometrically as a cylinder centered around
the chaotic state $\rho = \frac{1}{4}\Eins$ (compare with figure \ref{rBer}).
The function $f$ is easier to evaluate than the exponent $r$ and will be used in some calculations.
In the main theorem of the next chapter it will be demonstrated that purification succeeds in the regime of
characteristic exponents $r > 2$.

\subsection{Asymptotic evolution of \Pn\, steps}\label{AsymPn}
The evaluation of the asymptotic evolution of \Pn\, steps turns out to be much more complicated than
the one of \Bn\, steps. For this purpose the following lemma is useful:
\begin{Lemma}[Properties of the binomial distribution]\label{VarChernoff}\hfill\\
  Let $p \in [1/2;1]$, $n \in \N$ be odd; in these cases the relation
  \begin{equation}
 f_n(p) := \sum_{k = 0}^{(n-1)/2} \binom{n}{k}\, p^k (1-p)^{n-k} = c(n,p)\,z^n 
  \end{equation}
is valid with $z := 2\sqrt{p(1-p)}$, where the image of the function $c(n,p)$ is given by the interval
$[0;1]$ and $c(n,p)$ decreases at most sub-exponentially for \nGr\, and for any $p \in [1/2;1]$.
\end{Lemma}
\Bew A proof of this lemma is given in \ref{AnhChau}.
\par Analogous to (\ref{xnyn}) the asymptotic evolution of the state $(a,b,c,d)$ of a qubit pair under
a \Pn\, step is given by
\begin{equation}
  (a',b', c', d') = \left(\frac{1}{2} - u_n + v_n + \eps_n, u_n - \eps_n,
    \frac{1}{2} - v_n - \eps_n, \eps_n\right) := P_n\Bigl[(a,b,c,d)\Bigr].
\end{equation}
Here, $u_n$ is the \emph{phase error rate} and $v_n$ is the \emph{inverse bit error rate}; the value $\eps_n$
specifies the correlation between bit and phase errors.
\par Using these definitions, the calculation of $u_n$ and $v_n$ is straightforward, whereas the calculation of
the correlation $\eps_n$ is rather involved.  For odd values of $n \in \N$ one obtains the relations
\begin{equation}\begin{array}{rlrlrlrl}\label{PnUngl}
  u_n &= \sum_{k = 0}^{(n-1)/2} \binom{n}{k} (a+c)^k (b+d)^{n-k}
        && \stackrel{\mathrm{Lemma \ref{VarChernoff}}}{\leq} \bigl[4(a+c)(b+d)\bigr]^{n/2},\\
  2v_n &= (a+b-c-d)^n && \quad\equiv F^n.
\end{array}\end{equation}
Using lemma \ref{VarChernoff}  we may also write $u_n = c(n,a+c)\,z^n$ for $z = 2\sqrt{(a+c)(b+d)}$. Similar
to the construction for \Bn\, steps, one can define an exponent $r_P$, which characterizes the asymptotic
evolution of \Pn\, in the sense that $z/F^{r_P} = 1$. This yields the relation
\begin{equation}
  r_P = \frac{\ln z}{\ln F} = \frac{\ln 2\sqrt{(a+c)(b+d)}}{\ln (a+b-c-d)}
      = \frac{1}{2} \frac{\ln 4(a+c)(b+d)}{\ln (a+b-c-d)}
\end{equation}
for the characteristic exponent $r_P$. In view of the relation
\begin{equation}\label{KonvBed}
  \frac{u_n}{(2v_n)^{r_P}} = \frac{c(n,a+c)\,z^n}{F^{r_P n}} = c(n,a+c) \left(\frac{z}{F^{r_P}}\right)^n.
\end{equation}
the quotient $u_n/(2v_n)^{r_P}$ converges to $+\infty$ for all exponents larger than $r_P$
because $c(n,a+c) \leq 1$ decreases at most sub-exponentially.
Furthermore, the bounds $z,\, F \leq 1$ imply the inequalities ($B$ and $P$ denote bit and phase error rate):
\begin{equation}\begin{array}{rlrlrlrl}
  r_P > 1 &\eqv \left(1/2-B\right)^2 + \left(1/2-P\right)^2
            > \left(1/2\right)^2 = 1/4,\label{rP1}\\
  r_P > 2 &\eqv (1-2B)^4 - 4P(1-P) > 0.
\end{array}\end{equation}

\subsection{Remarks}
Note that the \Pn\, step defines a mapping $\Pn : (B,P) \mapsto (B',P')$, if one ignores the correlation
between bit and phase errors. In particular, a possible statistical independence of bit and phase errors,
i.\,e. the validity of the relation $(b+d)(c+d) - d = 0$, is invariant under \Pn\, steps but not under
\Bn\, steps. The following lemma is of some interest:
\begin{Lemma}[Separability using \Pn\, steps]\label{LemmaSep}\hfill\\
Let $\rho = (a,b,c,d) \in \Sv$, $n \in \N$ be odd and $\rho' = (a',b',c',d') := \Pn(\rho)$; this implies
  \begin{enumerate}
    \item $\rho'$ is entangled, if and only if $a' > 1/2$ holds.
    \item If bit and phase error rate in $\rho$\, are statistically independent, then for sufficiently
      large $n$ the state $\rho'$ is separable, if and only if $r_P(\rho) < 1$ holds.
  \end{enumerate}
\end{Lemma}
\Bew For the proof of the first statement, it is sufficient to show that $b',c',d' \leq 1/2$.
From (\ref{PnUngl}) follows the inequality $B' = c' + d' = (1 - F^n)/2 < 1/2$ and because of $F > 0$
one gets $c',d' \leq 1/2$. $P' = b' + d'$ decreases monotonically in $n$, which implies the assertion.
\par Thus, for the proof of the second inequality one concentrates on the value of~$a'$. Statistical
independence of bit and phase errors implies $a' = 1 - P' - B' + B'P'$; using the notation $c(n) := c(n,a+c)$
yields $a' = 1 - c(n)z^n - \left(1/2 - F^n/2\right) + c(n)z^n\left(1/2 - F^n/2\right)$ and
$a' \leq 1/2 \eqv \bigl(1-c(n)z^n\bigr)F^n \leq c(n)z^n$. Therefore, for a resulting separable state for
\nGr, $F^n \leq c(n)z^n$ is sufficient. Because $c(n)$ decreases at most sub-exponentially, $F < z$,
i.\,e. $r_P < 1$ is sufficient. On the contrary, if $r_P \geq 1$, i.\,e. $F \geq z$, the assertion follows
by a similar reasoning.

\section{The criterion for asymptotic correctability (main theorem)}\label{KorrKrit}
In this section the question of asymptotic correctability of Bell-diagonal quantum states is addressed
from a more general point of view. In particular, our main theorem is stated and proved which relates
the asymptotic correctability of a large class of general entanglement purification protocols to the
characteristic exponents determining the scaling of their resulting bit and phase errors. The general
entanglement purification protocols of this class are supposed to consist of arbitrary sequences of basic
steps which involve classical one- and/or two communication between Alice and Bob until the Shannon bound
is reached. Subsequently these steps are supposed to be completed by a CSS-based purification protocol,
which involves classical one-way communication. This main theorem will be specialized to sequences
of \Bn\, and \Pn\, steps in the next section.

Let us start by defining the notion of \emph{asymptotic correctability}:
\begin{Definition}[Asymptotic correctability]\hfill\\
  Let $\rho = (a,b,c,d) \in \Sv$ and $(S_n)_{n \in \N}$ be a sequence of possible steps in an
  entanglement purification protocol. The state $\rho$ is called \emph{asymptotically \Sn-correctable}
  under this sequence, if there exists an $N_0 \in \N$, such that for all $n \in \N$, $n \geq N_0$ the
  inequality $\AsymCSS\bigl[S_n(\rho)\bigr] := 1 - H(B) - H(P) > 0$ holds, where $B$ and $P$ denote
  bit and phase error rate of the resulting state $S_n(\rho)$ after the use of that step.
\end{Definition}
Here, $H(\xi) := -\xi\,\ld\,\xi - (1-\xi)\,\ld\,(1-\xi)$ is the binary Shannon entropy and the function
$\AsymCSS$\, denotes the Shannon bound, i.\,e. the minimum rate of an asymmetric CSS code \cite{CSS1,CSS2}.
If $\AsymCSS(\rho)$ is positive  the state $\rho$ can be corrected by some CSS code, i.\,e.
by one-way classical communication.
Important special cases are $(S_n)_{n \in \N} \in \Mg{(\Bn)_{n \in \N}, (P_{2n+1})_{n \in \N_0}}$.
Note that asymptotic correctability implies correctability, but not vice versa, in general.
\par Using the notation of (\ref{xnyn}) for the state of a qubit pair after application of an arbitrary
\Sn\, step, i.\,e. $B\to x_n$ and $P\to 1/2 - y_n$, one obtains
\begin{equation}\label{AsymTaylor}
  \AsymCSS(x_n, 1/2-y_n) = -H(x_n) + (\ln 2)^{-1}\left[2y_n\,\at(2y_n) + \frac{1}{2}\ln(1-4y_n^2)\right].
\end{equation}
Because of the symmetry of $\AsymCSS$, this is also valid for the case, where $P \to x_n$ and
$B \to 1/2-y_n$. Dropping positive terms in the (partial) Taylor series expansion of (\ref{AsymTaylor}) 
one obtains the lower bound
\begin{equation}
  \AsymCSS(x_n, 1/2-y_n) \geq A(x_n,y_n) := (\ln 2)^{-1}\left[x_n \ln x_n - x_n + 2y_n^2\right]
\end{equation}
for $0 \leq x_n \leq 1/2$ and $0 \leq y_n \leq 1/2$.
\par Obviously, one can define an \emph{asymptotic \Sn-correction} purification protocol in the following
way: Alice and Bob determine the smallest $n \in \N$, such that $S_n(\rho)$ can be corrected by some
asymmetric CSS code, apply \Sn, and use an appropriate CSS code to obtain a purified final state. 
In the case of \Bn\, and \Pn\, steps smaller values of $n$ usually result in higher key generation rates,
both in the two-way part of the protocol and in the CSS part.
\par Finally, it should be noted that the condition $\AsymCSS(\rho) > 0$ is only sufficient, but not
necessary for the existence of asymmetric CSS codes which are capable of purifying a quantum state.
If this condition is violated, there may also exist applicable
CSS codes, but this cannot be guaranteed in general.
\par After these introductory remarks let us state and prove now the following main theorem:
\begin{Satz}[Main theorem]\label{AsymKorr}\hfill\\
  Let $\rho = (a,b,c,d) \in \Sv$ and $(S_n)_{n \in \N}$ be a sequence of possible steps in an
  entanglement purification protocol. Furthermore, let
  \begin{equation*}
    (x_n,y_n) = (B, 1/2-P) \quad\mathit{or}\quad (x_n,y_n) = (P, 1/2-B)
  \end{equation*}
  after application of an \Sn\, step, and let $(S_n)_{n \in \N}$ be a sequence of such steps,
  such that $\lim_\nGr x_n~=~ 0$ holds. Finally, let
  \begin{equation}
    r_{\sup} := \sup\Mge{r\in\R}{\sup\Mge{x_n/y_n^r}{n\in\N} < \infty}.
  \end{equation}
  Then, $\rho$\, is asymptotically \Sn-correctable, if $r_{\sup} > 2$ holds. Furthermore,
  if $\rho$\, is asymptotically \Sn-correctable, then $r_{\sup} \geq 2$.
\end{Satz}
\Bew First part ($r_{\sup} > 2$ is sufficient): If $r_{\sup} > 2$, one can find an exponent $r > 2$
and a value $c > 0$, such that $x_n \leq cy_n^r$ for all $n \in \N$. The function $A(x,y)$ is used to
minorize $\AsymCSS(x,1/2-y)$. As a consequence the worst case with the maximum possible error rates is given
by $x_n = cy_n^r$. This implies
\begin{equation}\begin{array}{rlrlrlrl}
  (\ln 2\cdot A)(x_n,y_n) &= cy_n^r \ln(cy_n^r) - cy_n^r + 2y_n^2 > 0\\
    &\eqv \frac{c}{2} y_n^{r-2} \left[(\ln c + 1) + r \ln y_n\right] + 1 > 0.
\end{array}\end{equation}
Because $x_n$ tends to zero in the limit $n\to \infty$, also $y_n$ does so. Therefore, the first term of
the latter inequality becomes arbitrarily small due to $\lim_\nGr y_n^{r-2} \ln y_n = 0$. Thus,  we obtain
the required result, namely that $\AsymCSS(x_n,1/2-y_n) > 0$ for large $n$.
\par Second part ($r_{\sup} \geq 2$ is necessary): The condition $r_{\sup} < 2$ implies that
$\sup\Mge{x_n/y_n^2}{n\in\N} = \infty$, i.\,e. there exists at least a subsequence, for which
$c := \inf\Mge{x_n/y_n^2}{n\in\N} > 0$ holds. From the Shannon bound it is obvious, that for guaranteeing
correctability, $x_n$ should be as small and $y_n$ as large as possible. Therefore, in view of the
conditions of the theorem the best case is given by a subsequence with $x_n = cy_n^2$. Using
relation (\ref{AsymTaylor}) and the elementary properties
\begin{eqnarray}
  (\mathrm{d}/\mathrm{d}y) \left[2y\,\at(2y)\right. &+& \left.\ln(1-4y^2)/2\right] = 2\,\at(2y),\nonumber\\
  (\mathrm{d}/\mathrm{d}y) \bigl[2\,\at(2y)\bigr] &=& 4 / (4-y^2),\nonumber\\
  (\mathrm{d}/\mathrm{d}y) \left[-\ln 2\,H(cy^2) \right] &=& 2cy\,\ln\left(cy^2/(1-cy^2)\right),\nonumber\\
  (\mathrm{d}^2/\mathrm{d}y^2) \left[-\ln 2\,H(cy^2) \right] 
    &=& 2c \left[\ln \left(cy^2/(1-cy^2)\right) - 2/(cy^2-1) \right],
\end{eqnarray}
one therefore notices  
\begin{eqnarray}
  \lim_\nGr\AsymCSS(cy^2_n,1/2-y_n) &=& 0,\nonumber\\
  \lim_\nGr\frac{\mathrm{d}}{\mathrm{d}y}\AsymCSS(cy^2,1/2-y)\mid_{y=y_n} &=& 0,\nonumber\\
  \frac{\mathrm{d}^2}{\mathrm{d}y^2}\AsymCSS(cy^2,1/2-y)\mid_{y=y_n} &<& 0~~{\rm for}~~y_n \to 0.
\end{eqnarray}
Thus, the state is not asymptotically \Sn-correctable and the assertion is proved.
\par In particular, the special case $(S_n)_{n \in \N} \in \Mg{(\Bn)_{n \in \N}, (P_{2n+1})_{n \in \N_0}}$
yields
\begin{Korollar}[Asymptotic \Bn- and \Pn-correctability]\label{BnPnKorr}\hfill\\
  For $\rho \in \Sv$ the following statements are true:
  \begin{equation}\begin{array}{rlrlrlrl}
    \rho \mathrm{\,\,is\,asymptotically\,}\Bn\,\mathrm{correctable}
      &\eqv r(\rho) = \frac{\ln\frac{a+b}{c+d}}{\ln \frac{a+b}{a-b}} > 2,\\
    \rho \mathrm{\,\,is\,asymptotically\,}\Pn\,\mathrm{correctable}
      &\impl r_P(\rho) = \frac{\ln 4(a+c)(b+d)}{2\ln (a+b-c-d)} > 2.
  \end{array}\end{equation}
\end{Korollar}
\Bew This assertion follows immediately from theorem \ref{AsymKorr} and the basic properties of 
\Bn\, and \Pn\, steps discussed in sections \ref{AsymBn} and \ref{AsymPn}. The equivalence in the
case of asymptotic \Bn-correctability results from the fact that for $r = 2$ the equation
$\lim_\nGr x_n/y_n^2 = 4$ holds (see section \ref{AsymBn}); this implies $\inf\Mge{x_n/y_n^2}{n \in \N} > 0$
and the assertion follows as in the proof of theorem \ref{AsymKorr}.

\section{Asymptotic correctability using \Bn\, and \Pn\, steps}
In this section it is analyzed for which qubit-pair states $(a,b,c,d)$ a purification based on \Bn\,
and \Pn\, steps and asymmetric CSS codes fulfilling the Shannon bound is possible according to the main
theorem of the previous section. It is shown that bit-error correcting \Bn\, steps alone are already able
to guarantee security of the BB84 protocol and the six-state protocol up to maximum bit error rates of
magnitudes $1/5$ and $1/2-1/(2\sqrt{5})$, respectively. Furthermore, numerical evidence is provided that
even arbitrary sequences of phase-error correcting \Pn\, steps cannot improve on these bounds.
Based on this evidence the maximum possible bit error rates which are tolerable in the BB84 protocol
and in the six-state protocol are given by $1/5$ and $1/2-1/(2\sqrt{5})$, provided error correction and
privacy amplification are based on arbitrary sequences of \Bn\, and \Pn\, steps and the use of CSS codes.

\subsection{Reduction to the use of the exponent $r$}
So far we have concentrated on three possibilities for purifying a given Bell-diagonal quantum state.
A quantum cryptographic protocol can be made secure, if it produces states with $\AsymCSS(\rho)>0$,
$r(\rho) \equiv \frac{\ln\frac{a+b}{c+d}}{\ln \frac{a+b}{a-b}} > 2$ or
$r_P(\rho) \equiv \frac{\ln 4(a+c)(b+d)}{2\ln (a+b-c-d)}> 2$ and possibly in the case $r_P(\rho) = 2$.
As can be seen from the following theorem these conditions are not independent:
\begin{Satz}[Reduction to the characteristic exponent $r$]\label{rPr}\hfill\\
  Let $\rho = (a,b,c,d) \in \overline{\Sv}$. Then,
  \begin{equation}
    \AsymCSS(\rho) > 0 \impl r_P(\rho) > 1 \impl r(\rho) > 2.
  \end{equation}
  In particular, $r_P(\rho) \geq 2 \impl r(\rho) > 2$.
\end{Satz}
\Bew A detailed proof is given in \ref{Anh-rRed}.
\par It should be noted that for any state $\rho \in \Sbd$, the value of $r(\rho)$ is invariant
with respect to \Bn\, steps, because from (\ref{BnA}) one obtains immediately the relation
$r\bigl[\Bn(\rho)\bigr] = r(\rho)$.

\subsection{Limits for the  maximum tolerable error rate}
Theorem \ref{rPr} shows that it is sufficient to consider the characteristic exponent $r$\, for determining
the correctability using \Bn\, and \Pn\, steps and asymmetric CSS codes (using the Shannon bound).
According to this theorem the only possibility to purify states with $r \leq 2$ is to apply \Pn\, steps,
which may possibly yield states with $r > 2$.
If this is not possible, the asymptotic \Bn-correction is already optimal with \mbox{respect} to the
maximum tolerable error rate in our model. The following conjecture indeed suggests that asymptotic
\Bn\, correction is optimal:
\begin{Vermutung}[Optimality of the asymptotic \Bn\, correction]\label{OptBn}\hfill\\
  Let $\rho = (a,b,c,d) \in \overline{\Sv}$ with $r(\rho) \leq 2$. Then, for all odd $n \in \N$
  \begin{equation}
    r\bigl[P_n(\rho)\bigr] \leq 2.
  \end{equation}
\end{Vermutung}
\par
\begin{figure}[t]\begin{center}
  \includegraphics[height=140pt,width=150pt]{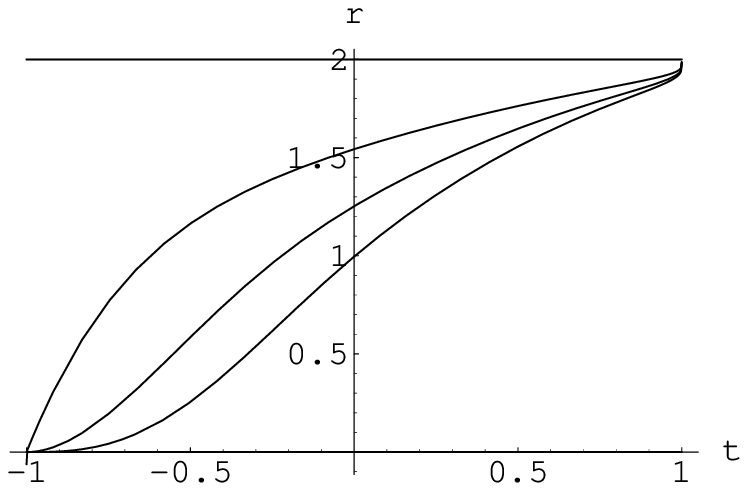} \qquad
  \includegraphics[height=140pt,width=150pt]{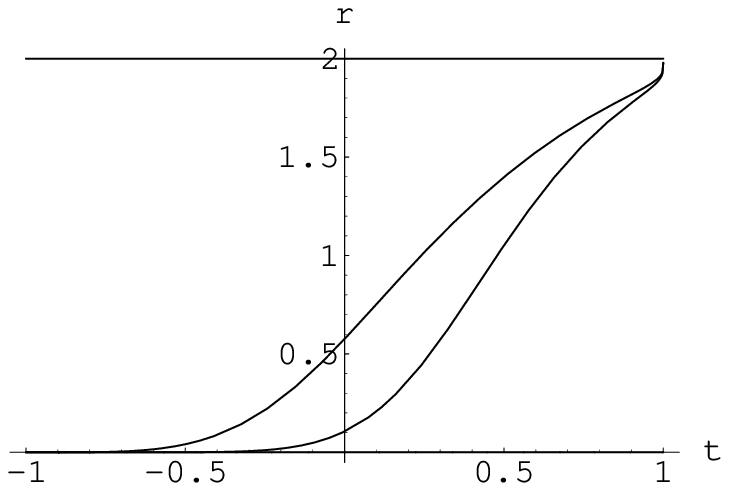}
  \caption{The function $r\bigl[P_n\bigl(K(t)\bigr)\bigr]$ for $n \in \Mg{3,5,7}$ (left) and
    $n \in \Mg{11,21}$ (right).} 
 \label{rPn}
\end{center}\end{figure}

 The subsequent lemmata show that for a proof of this conjecture it is sufficient to prove it
on a certain subset of states (compare with figure \ref{rBer}). But this turns out to be difficult and an
analytical proof is not known. However, as demonstrated below numerical
results (compare with figure \ref{rPn}) and plausibility arguments are in favour of the validity of this
conjecture.

For the formulation of these lemmata it is convenient to parameterize the set \Sv\, by
\begin{equation}
  Z(a,b;z) := \bigl(a,b, z(1-a-b), (1-z)(1-a-b)\bigr) \in \oSv
\end{equation}
with $a \geq 1/2$, $b \geq 0$, $a+b \leq 1$ and $z \in [0;1]$. 
It is useful to visualize these lemmata with the help of figure \ref{rBer}. The function $f$ introduced in
(\ref{fFkt}) will be used frequently.
\begin{Lemma}[Concerning the diagonals in figure \ref{rBer}]\label{Diag}\hfill\\
  Let $a, b, z, z', \delta \in [0;1]$ be chosen in such a way that $Z(a,b;z), Z(a-\delta,b+\delta;z')
  \in \oSv$. Then, $r\left[Z(a,b;z)\right] \leq 2 \impl r\left[Z(a-\delta,b+\delta;z')\right] \leq 2$.
\end{Lemma}
\Bew By (\ref{fFkt}), $r \leq 2 \eqv f(a,b) \leq 0$; thus, $z$ and $z'$ are unnecessary and one can calculate
$f(a-\delta,b+\delta) = f(a,b) + 2\delta(-a+b+\delta)$. The first expression is negative by assumption,
the factor $2\delta$ is non-negative. Using $Z(a-\delta,b+\delta;z') \in \oSv$, one finds $a - \delta \geq 1/2$
and therefore $\delta \leq a - 1/2 \leq a - b$, which implies the assertion.
\begin{Lemma}[First reduction to states with $d=0$]\label{LemmaRed1}\hfill\\
  Let $a, b \in [0;1]$ be chosen in such a way that $Z(a,b;1) \in \oSv$ and $f(a,b) \leq 0$, and let
  $n \in \N$ be odd and $z \in [0;1]$. Then, $r\bigl[P_n\bigl(Z(a,b;1)\bigr)\bigr] \leq 2
  \impl r\bigl[P_n\bigl(Z(a,b;z)\bigr)\bigr] \leq 2$.
\end{Lemma}
\Bew Let $\rho = Z(a,b;z) \in \Sv$. The \Pn\, step can be viewed as a mapping from old to new bit and
phase error rates, i.\,e. $\Pn : (B,P) \mapsto (B',P')$. In view of $B = c+d$ and $B' = c'+d'$ the bit error
rates do not depend on $z$. In figure \ref{rBer} a variation of $z$ results in a variation on the
diagonal $a'+b' = const.$ By the evolution (\ref{PnTransf}) one notes that the fidelity $a'$ becomes
larger, if the initial phase error rate gets small (proof in \ref{Red1}). Lemma \ref{Diag} now
implies the assertion.
\par Because of this, it is sufficient to consider the best case, i.\,e. $z = 1$ or $d = 0$.
\begin{Lemma}[Second reduction of the parameter space]\label{LemmaRed2}\hfill\\
  Let $a, b, \eps \in [0;1]$ be chosen in such a way that $Z(a,b;1), Z(a-\eps,b+\eps;1) \in \oSv$, and let
  $n \in \N$ be odd. Then,
  $r\bigl[P_n\bigl(Z(a,b;1)\bigr)\bigr] \leq 2 \impl r\bigl[P_n\bigl(Z(a-\eps,b+\eps;1)\bigr)\bigr] \leq 2$.
\end{Lemma}
\Bew The bit error rate $B = c+d$\, before and thus after a \Pn\, step does not depend on $\eps$. Using
lemma \ref{Diag}, in the best case the fidelity $a'$ is maximal after performing a \Pn\, step; as shown
in the \ref{Red2} this is the case for $\eps = 0$.
\par Because of the lemmata \ref{LemmaRed1} and \ref{LemmaRed2} the assertion from conjecture \ref{OptBn} has
to be shown only on a certain subset,  which can be parameterized by the function \Fkt{K}{[-1;+1]}{\oSv} with
\begin{equation}\label{Kante}
  K(t) := Z\left(1/4 + (2\sqrt{2})^{-1} \cos (\pi t / 4), 1/4 + (2\sqrt{2})^{-1} \sin (\pi t / 4); 1\right).
\end{equation}
This subset corresponds to the border of the black circle of figure \ref{rBer}.
Figure \ref{rPn} demonstrates graphically the validity of the claim for the first few values of $n$.
The curves of figure \ref{rPn} even seem to imply that $r$ tends to zero for large values of $n$. By Lemma
\ref{LemmaSep} it also appears that the states become separable and thus non-correctable for large values
of $n$.
\par Provided conjecture \ref{OptBn} is correct the following conjecture can be proven:
\begin{Vermutung}[Correctability by using \Bn\, and \Pn\, steps]\label{HptSatz}\hfill\\
  For $\rho = (a,b,c,d) \in \Sv$ the following statements are equivalent:
  \begin{enumerate}
    \item $r(\rho) > 2$ (or equivalent $f(a,b) > 0$ by (\ref{fFkt}));
    \item $\rho$\, is asymptotically \Bn-correctable;
    \item There exists a sequence of \Bn\, and \Pn\, steps, such that after performing this sequence
      the resulting state $\rho'$\, fulfills the inequality $\AsymCSS(\rho') > 0$.
  \end{enumerate}
\end{Vermutung}
\Bew The equivalence of the first two statements was shown in corollary \ref{BnPnKorr} on page
\pageref{BnPnKorr}; that the second statement implies the third one is trivial, and that the third
one implies the first follows from theorem \ref{rPr} and conjecture \ref{OptBn} via contraposition.

\subsection{Values of the maximum tolerable error rate}
Using the criterion derived in the previous sections, one can calculate the maximum tolerable error rate
for the BB84 and the six-state protocol  assuming the model considered there. In case of the
six-state protocol $b = c = d$ holds \cite{GL}; thus, one only has to consider the so-called Werner states.
Using the notation
\begin{equation}\begin{array}{rlrlrlrl}
  \W(F)    &:= \left(F,\frac{1-F}{3}, \frac{1-F}{3}, \frac{1-F}{3}\right), \qquad
  \BB(F) &:= \left(F,\frac{1-F}{2}, \frac{1-F}{2}, 0\right),
\end{array}\end{equation}
one calculates for the six-state protocol
\begin{equation}\begin{array}{rlrlrlrl}
  &r\bigl[\W(F)\bigr] > 2 &&\eqv F > \bigl(5 + 3\sqrt{5}\bigr)/20 &&\approx 0.585410\\
                         &&&\eqv B < 1/2 - 1/(2\sqrt{5}) &&\approx 27.6393\,\%.
\end{array}\end{equation}
For the BB84 protocol one can in principle use similar reasoning as the one by Gottesman-Lo \cite{GL},
but the statement that the $\BB(F)$ state is the worst case for fixed bit error rate $B$ can be proved
much easier now. As before $B = P = b+d = c+d$ and thus $b = c$ hold; using a suitable parameter
$\delta \in [0;B]$, one can rewrite the state as
\begin{equation}
  \rho = (1 - 2B + \delta, B - \delta , B - \delta, \delta).
\end{equation}
By (\ref{fFkt}) it follows that $f(\rho) = 2\delta^2 + (2-6B)\delta + (1/2 - 7B/2 + 5B^2)$ and derivation
with respect to $\delta$ yields $4 \delta^2 + (2-6B) \geq 0$, if $B \leq \dd\,\%$. Therefore, $f$ increases
monotonically with respect to $\delta$ and the worst case possible is $\delta = 0$, i.\,e. the $\BB$ state
defined above. In this case, it follows
\begin{equation}\begin{array}{rl}
  r\bigl[\BB(F)\bigr] > 2 &\eqv F > 3/5 = 0.600000\\
                          &\eqv B < 1/5 = 20.0000\,\%.
\end{array}\end{equation}
These maximum tolerable error rates coincide exactly with the ones given by Chau \cite{Ch}.

\section{Conclusions}
We analyzed general entanglement purification protocols which imply the security of any quantum key
distribution protocol whose security analysis can be reduced to the purification of Bell-diagonal states.
These entanglement purification protocols are supposed to consist of arbitrary sequences of basic steps
involving classical one- and/or two-way communication between Alice and Bob until the Shannon bound
guarantees a successful completion of the entanglement purification on the basis of an appropriate
CSS encoding and classical one-way communication. As a main result a necessary and a sufficient condition
on asymptotic correctability of Bell-diagonal qubit-pair states were presented relating the success of
such general entanglement purification protocol to the magnitude of a characteristic exponent, which governs
the scaling between bit and phase errors. 
Applying this theorem to entanglement purification protocols of the Gottesman-Lo type we demonstrated that
in the cases of the BB84 and six-state quantum cryptographic protocols secret keys can be generated 
even without any phase-error correcting steps of the Gottesman-Lo type up to the already known bit error
rates of $1/5 = 20\% $ and $1/2-1/(2\sqrt{5})\approx 27.6393\, \%$. 
Furthermore, numerical evidence was provided that also the inclusion of additional arbitrary sequences
of phase-error correcting steps cannot improve on these particular bounds.

\subsection*{Acknowledgements}
This work is supported by the EU within the IP SECOQC. Informative discussions with 
A. Khalique, N. L\"utkenhaus and G. Nikolopoulos are acknowledged. K. Ranade is supported by
a graduate-student scholarship of the Technische Universit\"at Darmstadt.

\appendix

\section{Evolution using \Bn\, and \Pn\, steps}

\subsection{Evolution using \Bn\, steps}\label{AnhBn}
On two possibly different states $\rho = (a,b,c,d) \in \Sbd$ and $\sigma = (p,q,r,s) \in \Sbd$ a $B_2$\,
step is applied. After measuring and discarding the second qubit pair, the reduced density matrix of the
first pair reads
\begin{equation}\label{BAsym}
  \rho' = \left(\frac{ap+bq}{N}, \frac{bp+aq}{N}, \frac{cr+ds}{N}, \frac{dr+cs}{N}\right),
\end{equation}
where $N = (a+b)(p+q) + (c+d)(r+s)$ is the normalization constant.
\par The proof of formulae (\ref{BnA}) will be done by induction similar to the one in
\cite{Ch}. In a \Bn\, step the $B_2$\, step is used $(n-1)$ times, where $\rho$ is the first pair
and $\sigma$ is a new pair every time, i.\,e. $\rho = B_k\left[(a,b,c,d)\right]$ and $\sigma = (a,b,c,d)$.
One notes, that the case $n = 1$ is trivial and $n = 2$ is the starting point of the induction. One
now assumes that formulae (\ref{BnA}) are  valid for a fixed $n \in \N$. By using (\ref{BAsym})
one calculates for $(a',b',c',d') := B_{n+1}\bigl[(a,b,c,d)\bigr]$
\begin{equation}\begin{array}{rlrlrlrl}
   a' &= \bigl[(a+b)^{n+1}+(a-b)^{n+1}\bigr]/2N' \quad
   b' &= \bigl[(a+b)^{n+1}-(a-b)^{n+1}\bigr]/2N'\\
   c' &= \bigl[(c+d)^{n+1}+(c-d)^{n+1}\bigr]/2N' \quad
   d' &= \bigl[(c+d)^{n+1}-(c-d)^{n+1}\bigr]/2N'\\
\end{array}\end{equation}
where $N' = \bigl[(a+b)^{n+1} + (c+d)^{n+1}\bigr]$ is the new normalization constant.

\subsection{Evolution using \Pn\, steps}\label{AnhPn}
The evolution of a state by applying \Pn\, steps is more complicated than the one by applying
\Bn\, steps. An analytical expression can be given by listing all possible combinations  of Bell states,
calculating the resulting Bell state systematically (by phase majority and bit parity) and adding them up
according to their probability; for $\Pn\bigl[(a,b,c,d)\bigr]$ it follows:
\begin{equation}\label{EntwPn}
\sum_{(A,B,C,D) \in X_n}
    M(A,B,C,D) \bigl(a^A b^B c^C d^D, a^B b^A c^D d^C, a^C b^D c^A d^B, a^D b^C c^B d^A\bigr).
\end{equation}
Here, $M(A,B,C,D) := (A+B+C+D)! / (A!\,B!\,C!\,D!)$ is a multinomial coefficient and
$X_n := \Mge{(A,B,C,D) \in \N_0^4}{A+B+C+D = n, A+C > B+D, A+B \mathrm{\,odd}}$.

\section{Remarks to some theorems}

\subsection{Proof of lemma \ref{VarChernoff}}\label{AnhChau}
The idea of lemma \ref{VarChernoff} (page \pageref{VarChernoff}) is to determine the exponential evolution of
$f_n(p)$ and to absorb it into the value of $z^n$. Therefore, the appropriate value is $z(p) = \lim_\nGr
\sqrt[n]{f_n(p)}$. In particular, $z(1/2) = 1$ and $z(1) = 0$.
For the remaining cases $p \in (1/2;1)$, one uses only the last term in the expression for $f_{2n+1}(p)$,
which leads to
\begin{equation}
  f_{2n+1}(p) = \sum_{k=0}^{n} \binom{2n+1}{k}\,p^k\,(1-p)^{2n+1-k} \geq \binom{2n+1}{n}\,p^n\,(1-p)^{n+1}.
\end{equation}
The Stirling formula \cite{AS} $n^ne^{-n}\sqrt{2\pi n} \leq n! \leq n^ne^{-n}\sqrt{2\pi n}\,
e^{1/12n}$ yields
\begin{equation}
  \frac{n+1}{2n+1} \cdot \binom{2n+1}{n} = \frac{(2n)!}{(n!)^2} \geq
  \frac{(2n)^{2n}\,e^{-2n}\,\sqrt{2\pi (2n)}}{n^{2n}\,e^{-2n}\,2\pi n\, e^{1/6n}}
  = 2^{2n}\,\frac{e^{-1/6n}}{\sqrt{\pi n}}.
\end{equation}
Thus, $\binom{2n+1}{n} \geq 2^{2n+1} h(n)$ with $h(n) = e^{-1/6n} \bigl(1-\frac{1}{2(n+1)}\bigr)/\sqrt{\pi n}$
and therefore
\begin{equation}
  f_{2n+1}(p)^{\frac{1}{2n+1}} \geq 2\,h(n)^{\frac{1}{2n+1}}\, p^{\frac{1}{2+\frac{1}{n}}}\,
  (1-p)^{\frac{1}{2-\frac{1}{n+1}}} \stackrel{\nGr}{\longrightarrow} 2\sqrt{p(1-p)}.
\end{equation}
By this $z(p) \geq 2\sqrt{p(1-p)}$ was proved. The inequality $z(p) \leq 2\sqrt{p(1-p)} $ is a special case
of the Chernoff bound (cf. \cite{NC}, p. 154, (3.5)). 

\subsection{Proof of theorem \ref{rPr}}\label{Anh-rRed}
\begin{figure}[t]\begin{center}
  \includegraphics[height=150pt,width=150pt]{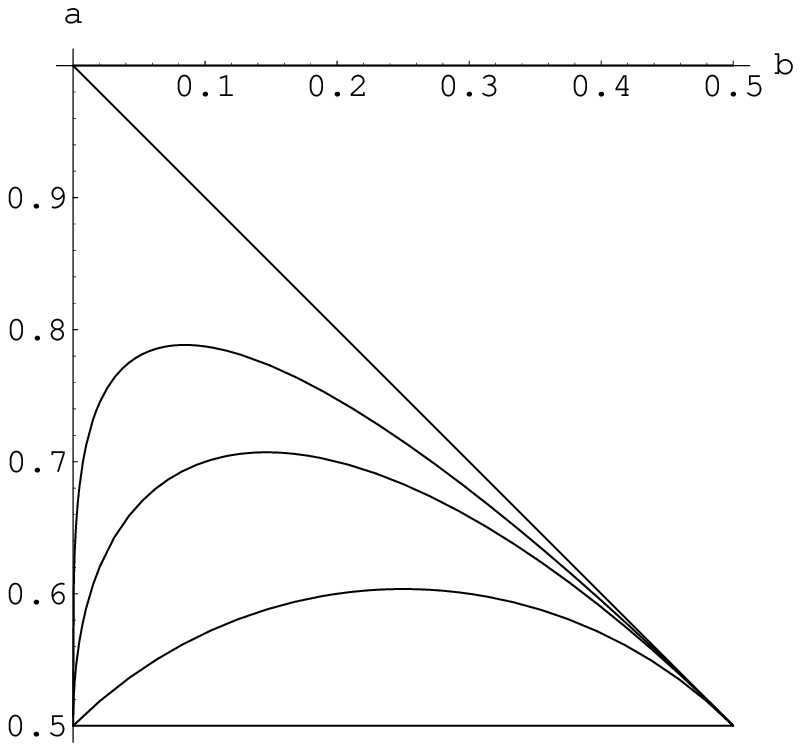} \qquad
  \includegraphics[height=155pt,width=155pt]{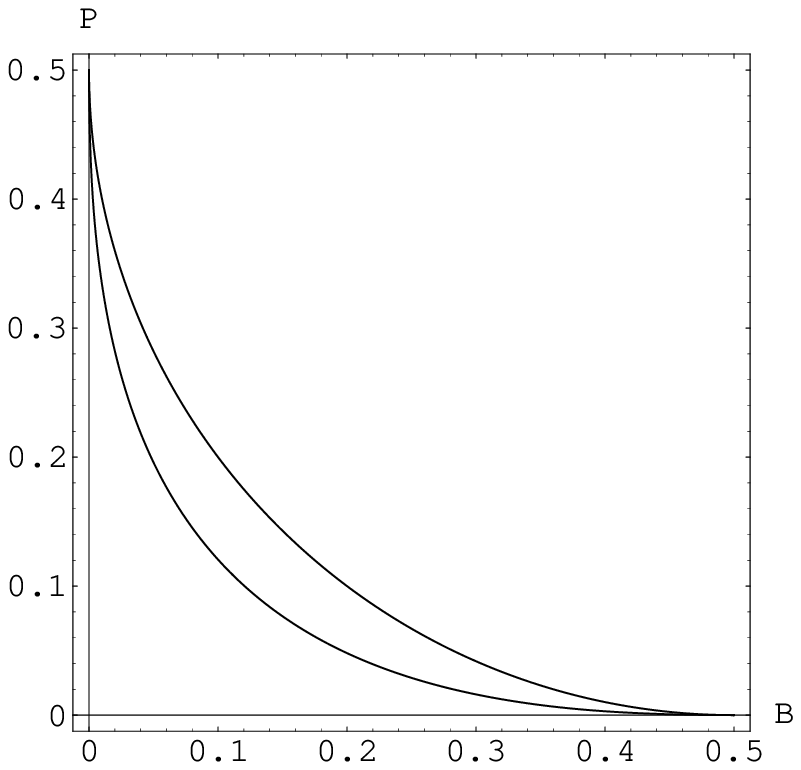}
  \caption{Left figure: minimum fidelity for $r > 2$, $r_P > 1$ and $r_P > 2$ (down to up);
     right figure: Lines for $\AsymCSS(B,P) = 0$ and $r_P = 1$}
 \label{rrPVgl}
\end{center}\end{figure}

\subsubsection{On the first implication ($\AsymCSS(\rho) > 0 \impl r_P > 1$)}
For the proof of the first implication, one notes that $\AsymCSS$ and $r_P$ can be considered as functions
of $B$ and $P$ and that $\AsymCSS(B_1,P_1) \geq \AsymCSS(B_2,P_2)$ holds, if $0 \leq B_1 \leq B_2 \leq 1/2$
and $0 \leq P_1 \leq P_2 \leq 1/2$. Because of (\ref{rP1}) it has to be shown that $\AsymCSS(B,P) \leq 0$
is true on the circular arc defined by $r_P = 1$ (see figure \ref{rrPVgl}), i.\,e. that
\begin{equation}
  h(t) := 1 - H\bigl[(\cos t)/2\bigr] - H\bigl[(\sin t)/2\bigr] \leq 0
\end{equation}
is valid for $t \in [0;\pi/2]$; by symmetry of the function, it is sufficient to show the property
for $t \in [0;\pi/4]$. Using $h(0) = 0$, it is further sufficient to show that $h'(t) \leq 0$
for $t \in [0;\pi/4]$, i.\,e.
\begin{equation}
  (\ln 4 \cdot h')(t) = \cos t \bigl[\ln \sin t - \ln (2-\sin t) \bigr] 
                      - \sin t \bigl[\ln \cos t - \ln (2-\cos t) \bigr] \leq 0.
\end{equation}
Rewriting this inequality yields $\sin t \bigl[\ln (2-\cos t) - \ln \cos t \bigr] \leq \cos t
\bigl[\ln (2-\sin t) - \ln \sin t \bigr]$ and because $t \in [0;\pi/4]$ implies $\cos t \geq \sin t \geq 0$,
it further only remains to show that
\begin{equation}
  h'_B(t) := \ln (2-\cos t) - \ln \cos t - \ln (2-\sin t) + \ln \sin t \leq 0
\end{equation}
is valid. By $h''_B(t) = (\sin t/(2 - \cos t)) + \tan t + (\cos t/(2 - \sin t)) + \cot t$, $h''_B(t) \geq 0$
for $t \in [0;\pi/4]$, and thus, $h'_B$ increases monotonically. Finally, $h'_B(\pi/4) = 0$, which implies
the assertion. 

\subsubsection{On the second implication ($r_P > 1 \impl r > 2$)}
The proof of the second implication can also be visualized by figure \ref{rrPVgl}. Plotting the minimum
fidelity $a \in [1/2;1]$, for which $r > 2$ is true, as a function of $b \in [0;1/2]$ results in
the function
\begin{equation}
  f_{r = 2}(b) := 1/4 + \sqrt{1/8 - \left(b - 1/4\right)^2}.
\end{equation}
Because $r_P$ depends upon the error rates $B$ and $P$, it is not directly possible, to plot the minimum
fidelity $a$ as a function of $b$. Assuming the best case (i.\,e. the smallest minimum fidelity possible),
one assumes the minimum phase error rate and therefore $d = 0$. In this case the limiting function is
\begin{equation}
  f_{r_P = 1}(b)  := 1 - b - (1/2 - \sqrt{b(1-b)}).
\end{equation}
For proving $f_{r_P = 1} \geq f_{r = 2}$ (see also figure \ref{rrPVgl}), let $\Delta(b) := f_{r_P = 1}(b)
- f_{r = 2}(b)$.
It has to be shown that $\Delta(b) \geq 0$ for $b \in [0;1/2]$. This function is continuous and by the
intermediate value theorem, it is sufficient to show that $b_1 = 0$ and $b_2 = 1/2$ are the only points
where it is zero and that there exists a point $b$ where $\Delta(b) > 0$. Repeated squaring of the
equation $\Delta(b) = 0$ yields a necessary condition for any zero of $\Delta$:
\begin{equation}
  5b^4-6b^3+9b^2/4-b/4 = 5b(b-1/5)(b-1/2)^2 = 0.
\end{equation}
The set of zeroes of the last equation is \Mg{0, 1/5, 1/2}. Because of $\Delta(0) = \Delta(1/2) = 0$
and $\Delta(1/5) = 1/10 > 0$, $\Delta$ is non-negative on the whole interval $[0;1/2]$. 

\subsection{Remarks to conjecture \ref{OptBn}}
Some details regarding lemmata \ref{LemmaRed1} and \ref{LemmaRed2} are given.
Before continuing, note the following lemma (the proof is trivial):
\begin{Lemma}[Monotonicity of the binomial distribution]\label{MonotBV}\hfill\\
  Let $n \in \N_0$ and $r \in \MgN{n}$. The function \Fkt{f}{[0;1]}{[0;1]}, which is defined by
    $f(x) := \sum_{k = 0}^r \binom{n}{k}\,x^k (1-x)^{n-k}$
  decreases monotonically in $x$.
\end{Lemma}

\subsubsection{On the first reduction}\label{Red1}
It remains to show, that $a'$ is maximal, if $z = 1$. By (\ref{EntwPn}) it follows using $\rho = Z(a,b;z)$
and $K := C+D$ and $(A,B,C,D) \in X_n$ that
$a' = \sum_{A,B} M(A,B,C+D,0)a^A b^B (1-a-b)^{C+D} \sum_{D = 0}^{D_{\max}} \binom{K}{D} z^{K-D} (1-z)^D$.
For $a'$ being maximal, it is sufficient that for all possible $A,B,K$ each term of the inner sum becomes
maximal. For fixed $A$ and $B$ the sum over $D$ is of such a form, that lemma \ref{MonotBV} can be applied,
i.\,e. $a'$ becomes maximal when $(1-z) = 0$ or $z = 1$ hold. 

\subsubsection{On the second reduction}\label{Red2}
The proof is similar to the previous one. Using $K := A + B$ yields $a' = \sum_C \binom{n}{C}
c^C \sum_{B=0}^{B_{\max}} \binom{K}{B} (a-\eps)^{K-B} (b+\eps)^B$.
As before the maximality of the inner sum is sufficient for the maximality of $a'$. If one divides
this by $(a+b)^K$, the assertion follows by lemma~\ref{MonotBV}.

\end{document}